\PassOptionsToPackage{unicode=true}{hyperref} 
\PassOptionsToPackage{hyphens}{url}
\PassOptionsToPackage{dvipsnames,svgnames*,x11names*}{xcolor}
\documentclass[10pt,]{article}

\usepackage{amssymb,amsmath}
\usepackage{ifxetex,ifluatex}
\usepackage{fixltx2e} 
\ifnum 0\ifxetex 1\fi\ifluatex 1\fi=0 
  \usepackage[T1]{fontenc}
  \usepackage[utf8]{inputenc}
  \usepackage{textcomp} 
\else 
  \usepackage{unicode-math}
  \defaultfontfeatures{Ligatures=TeX,Scale=MatchLowercase}
\fi
\IfFileExists{upquote.sty}{\usepackage{upquote}}{}
\IfFileExists{microtype.sty}{%
\usepackage[]{microtype}
\UseMicrotypeSet[protrusion]{basicmath} 
}{}
\IfFileExists{parskip.sty}{%
\usepackage{parskip}
}{
\setlength{\parindent}{0pt}
\setlength{\parskip}{6pt plus 2pt minus 1pt}
}
\usepackage{xcolor}
\usepackage{hyperref}
\hypersetup{
            pdftitle={OmniPilot: An Uncertainty-Aware LLM Inference Advisor for Heterogeneous GPU Clusters},
            pdfauthor={D. Balamurugan and Thomas W. Bush},
            colorlinks=true,
            linkcolor=Maroon,
            citecolor=Blue,
            urlcolor=Blue,
            breaklinks=true}
\usepackage[margin=1in]{geometry}
\usepackage{longtable,booktabs}
\IfFileExists{footnote.sty}{\usepackage{footnote}\makesavenoteenv{longtable}}{}
\providecommand{\tightlist}{%
  \setlength{\itemsep}{0pt}\setlength{\parskip}{0pt}}
\ifx\paragraph\undefined\else
\let\oldparagraph\paragraph
\renewcommand{\paragraph}[1]{\oldparagraph{#1}\mbox{}}
\fi
\ifx\subparagraph\undefined\else
\let\oldsubparagraph\subparagraph
\renewcommand{\subparagraph}[1]{\oldsubparagraph{#1}\mbox{}}
\fi

\makeatletter
\def\fps@figure{htbp}
\makeatother

\usepackage{booktabs}
\usepackage{cite}
\usepackage{graphicx}
\usepackage{float}
\usepackage{authblk}
\graphicspath{{figures/}}

\title{OmniPilot: An Uncertainty-Aware LLM Inference Advisor for
Heterogeneous GPU Clusters}

\author[1]{D. Balamurugan\thanks{Corresponding author: \texttt{bala\_desinghu@harvard.edu}}}
\author[1]{Thomas W. Bush}
\affil[1]{Kempner Institute for the Study of Natural and Artificial Intelligence, Harvard University}

\begin{document}
\maketitle

\begin{abstract}

Serving large language models (LLMs) on a shared, heterogeneous GPU cluster requires users and operators to select the GPU type, tensor-parallel degree, and precision before committing valuable node-hours. Making these choices is challenging because effective throughput, launch-success rates, and cluster demand and utilization continuously fluctuate. Furthermore, static configuration recipes miss critical interactions: quantization effects depend heavily on the model family, key-value cache pressure creates size-by-precision trade-offs, and failure rates vary by more than twofold across different tensor-parallel degrees. Additionally, cluster resources are frequently constrained by unpredictable hardware failures.  To address these challenges, we present \textbf{OmniPilot}, a launch advisor that predicts serving costs for feasible configurations and abstains when requests fall outside its measured support envelope. OmniPilot pairs a conformally calibrated quantile cost model (spanning eight serving targets) with an out-of-distribution (OOD) abstention layer. It ranks configurations using an economic utility metric calibrated to an operator's revealed preferences. In evaluations across 460 benchmark runs on A100, H100, and H200 hardware across four precisions, OmniPilot predicts aggregate throughput with a 6.2\% mean absolute percentage error (MAPE) and a log-space $R^2=0.92$. Compared to operator-relevant, model-free baselines, the advisor achieves 95\% top-1 accuracy with a mean utility regret of just 0.003. When tested on an OOD holdout of unsupported cells, prediction error climbs to 24–46\% and conformal intervals cover 0 of 5 points; however, the abstention layer successfully flags all five as low-confidence. Over time, these OOD scenarios will be integrated into the training dataset to continuously expand the advisor's support envelope. Training and recovery workloads remain out of scope for this work, but are slated for future research.

\smallskip
\noindent\textbf{Keywords:} LLM inference serving, GPU clusters, cost
modeling, conformal prediction, abstention, uncertainty quantification,
quantization, MLOps.
\end{abstract}

\hypertarget{introduction}{%
\subsection{1. Introduction}\label{introduction}}
  
\hypertarget{first-attempt-failures-on-shared-clusters-are-expensive}{%
\subsubsection{1.1 First-attempt failures on shared clusters are expensive}\label{first-attempt-failures-on-shared-clusters-are-expensive}}

Shared GPU clusters are highly contended, multi-tenant, and heterogeneous environments. Before an inference server can deliver its first token, an operator\footnote{Hereafter, we use the term \emph{operator} as a convenient shorthand to represent any entity submitting a workload, including individual end-users, system administrators, or delegated AI agents managing MLOps pipelines.} must select the underlying hardware configuration: the specific GPU type, the tensor-parallel (TP) degree, and the numerical precision. Making these decisions blindly is costly. Out of our 561 inference benchmark runs, 80 runs (14\%) ended in immediate failure due to hard hardware incompatibilities or out-of-memory resource exhaustion.

Static, one-size-fits-all deployment recipes are brittle in this complex decision space. Intricate hardware-software interactions create highly counterintuitive behaviors: a 4-bit quantized weight format can actually run slower than a 16-bit format on a given runtime, key-value (KV) cache pressure introduces dynamic size-by-precision trade-offs, and multi-GPU launch success rates degrade sharply as the TP degree increases. Because these low-level runtime nuances are entirely invisible to traditional cluster orchestration, operators require a pre-submission advisor. This system must provide performance predictions with calibrated uncertainty, alongside an explicit abstention signal to block unsupported or high-risk requests.

\textbf{Scope:} Motivated by our cluster job history data (§5.2) from Slurm and DCGM collector, which reveals that the vast majority of production inference jobs target a single node, this paper focuses exclusively on single-node vLLM {[}1{]} inference serving. Training workloads and fault recovery are deferred to future work (§5.5).

\hypertarget{the-gap}{%
\subsubsection{1.2 The gap}\label{the-gap}}

At the cluster orchestration layer, traditional job schedulers like Slurm merely match explicit user resource requests to the immediate, static state of the cluster. These schedulers operate with a critical blind spot: they possess no historical awareness regarding the behavioral nature of the specific workload, nor do they monitor wide-range telemetry signals such as Data Center GPU Manager (DCGM) metrics or complex network parameters.

Existing serving systems {[}1{]}--{[}5{]} primarily optimize execution \emph{within} a pre-selected configuration, focusing on KV-cache management, iteration-level scheduling, and prefill/decode balancing. These optimizations, performed by specialized inference engines like vLLM or TensorRT, are highly effective but fundamentally localized. While recent systems disaggregate serving phases or route traffic across heterogeneous GPUs {[}21{]}--{[}26{]}—with Helix {[}24{]} being the closest comparator—they typically route requests over a fixed, pre-provisioned fleet rather than dynamically selecting individual launch configurations. Furthermore, current quantization literature {[}6{]}--{[}8{]} emphasizes accuracy retention rather than the specific throughput and power envelopes that each precision yields on shared hardware.

This leaves a critical gap at the macro-level of multi-user, shared GPU clusters: finding optimal, pre-submission configurations to serve a massive volume of inference requests from diverse demands. Operators must determine the optimal GPU, precision, and TP degree \emph{before} committing valuable node-hours, complete with calibrated uncertainty. OmniPilot is designed specifically to fill this gap. By ingesting an expansive array of observable historical metrics alongside live cluster telemetry—including DCGM performance data and network constraints—OmniPilot selectively learns and predicts the specific demands of machine learning workloads. It acts as a cluster-scale launch planner that \emph{complements} the execution-level optimizations of inference engines. By handling the complex economic and placement decisions upfront, OmniPilot ensures that high-volume, multi-tenant inference workloads are placed on the most efficient configurations, allowing engines like vLLM to maximize performance once the job begins.

\hypertarget{contributions}{%
\subsubsection{1.3 Contributions}\label{contributions}}

\begin{enumerate}
\def\labelenumi{\arabic{enumi}.}
\item
  \textbf{A calibrated, abstention-aware launch advisor.} OmniPilot uses a
  gradient-boosted quantile cost model over eight serving targets, nine
  model families, five GPU kinds, and four quantizations. It combines
  conformal prediction intervals calibrated on held-out-cell residuals
  with an out-of-distribution abstention layer (§3.2). This combination
  jointly predicts throughput, TTFT, KV saturation, and power under
  quantization on shared-cluster hardware, while exposing calibrated
  intervals and explicit abstention.
\item
  \textbf{Decision-coupled evaluation against operator baselines.}
  Leakage-free GroupKFold backtests compare OmniPilot with model-free
  launch policies (§3.3, §4). OmniPilot reaches 95\% top-1 placement
  accuracy on the frozen 460-row evaluation and 92\% on the expanded
  dataset (Table~\ref{tab:baselines}), with mean utility regret more than
  an order of magnitude below size-threshold, H100/FP8-recipe, and
  throughput-max policies.
\item
  \textbf{A cluster-telemetry study with a negative result.} We mine
  500K Slurm jobs into per-configuration launch-success and demand
  priors (§5.2). The same harvest shows that scheduler metadata cannot
  support a cluster-wide throughput advisor because it records how a GPU
  was used, but not what workload ran on it.
\end{enumerate}

We also develop three supporting methods: a structurally validated
value-of-information benchmark gate (§5.4), a feature-before-data
methodology for new modeling axes (§3.2.1), and a regression-gated update
loop demonstrated on one promotion cycle (Appendix~C.2).

\begin{figure}[H]
\centering
\includegraphics[width=0.44\linewidth]{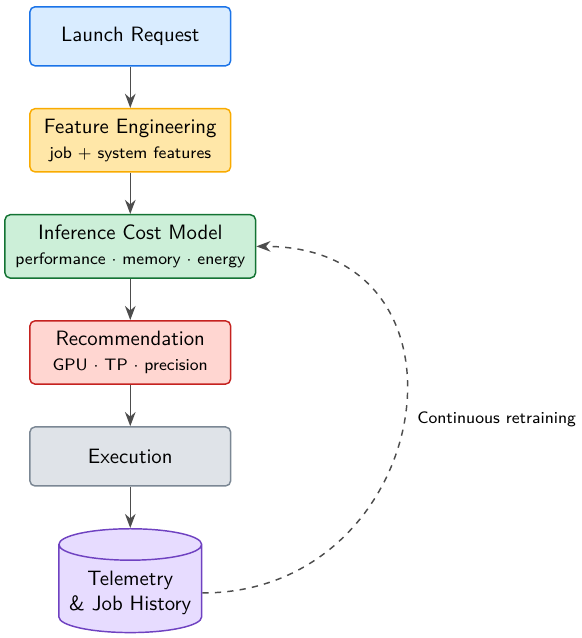}
\caption{\textbf{OmniPilot pipeline.}
When a launch request arrives, the cost model analyzes it to predict performance, memory use, and energy consumption, complete with a calibrated margin of error. Based on these predictions, the decision layer recommends the optimal GPU type, tensor-parallel degree, and precision. Once the job runs, its actual performance is logged and fed back into the model through an automated retraining loop to continuously improve accuracy. (Detailed descriptions of the system's internal mechanisms—including telemetry, calibration, and safety fallbacks—are provided in §3.1 and Appendix C.)
}
\label{fig:arch}
\end{figure}

\begin{center}\rule{0.5\linewidth}{0.4pt}\end{center}

\hypertarget{background-and-motivation}{%
\subsection{2. Background and
Motivation}\label{background-and-motivation}}

\hypertarget{the-cluster}{%
\subsubsection{2.1 The cluster}\label{the-cluster}}

Our deployment is the Kempner AI cluster {[}27{]}, a production shared
cluster scheduled by Slurm and one of the largest dedicated AI computing
clusters at an academic institution. It
contains A100 (40 GB), H100 (80 GB), and H200 (141 GB) GPUs. Memory
capacity interacts with model size and precision, which changes the
lowest-cost feasible GPU. Co-tenant jobs can also reduce achievable
throughput by consuming memory or bandwidth. Every result in this paper
uses single-node inference serving.

\hypertarget{the-inference-launch-decision-space}{%
\subsubsection{2.2 The inference launch decision
space}\label{the-inference-launch-decision-space}}

The \textbf{launch decision} takes a serving request and the current cluster state as input. This cluster state is dynamically constructed by fusing live scheduler metadata from Slurm with deep hardware telemetry, capturing both cluster-wide metrics (such as multi-tenant node contention and global queue depths) and job-specific metrics (such as runtime history and explicit resource requirements). The incoming request specifies the model, expected concurrency, target context length, and an optional service-level objective (SLO). 

\begin{figure}[H]
\centering
\includegraphics[width=0.7\linewidth]{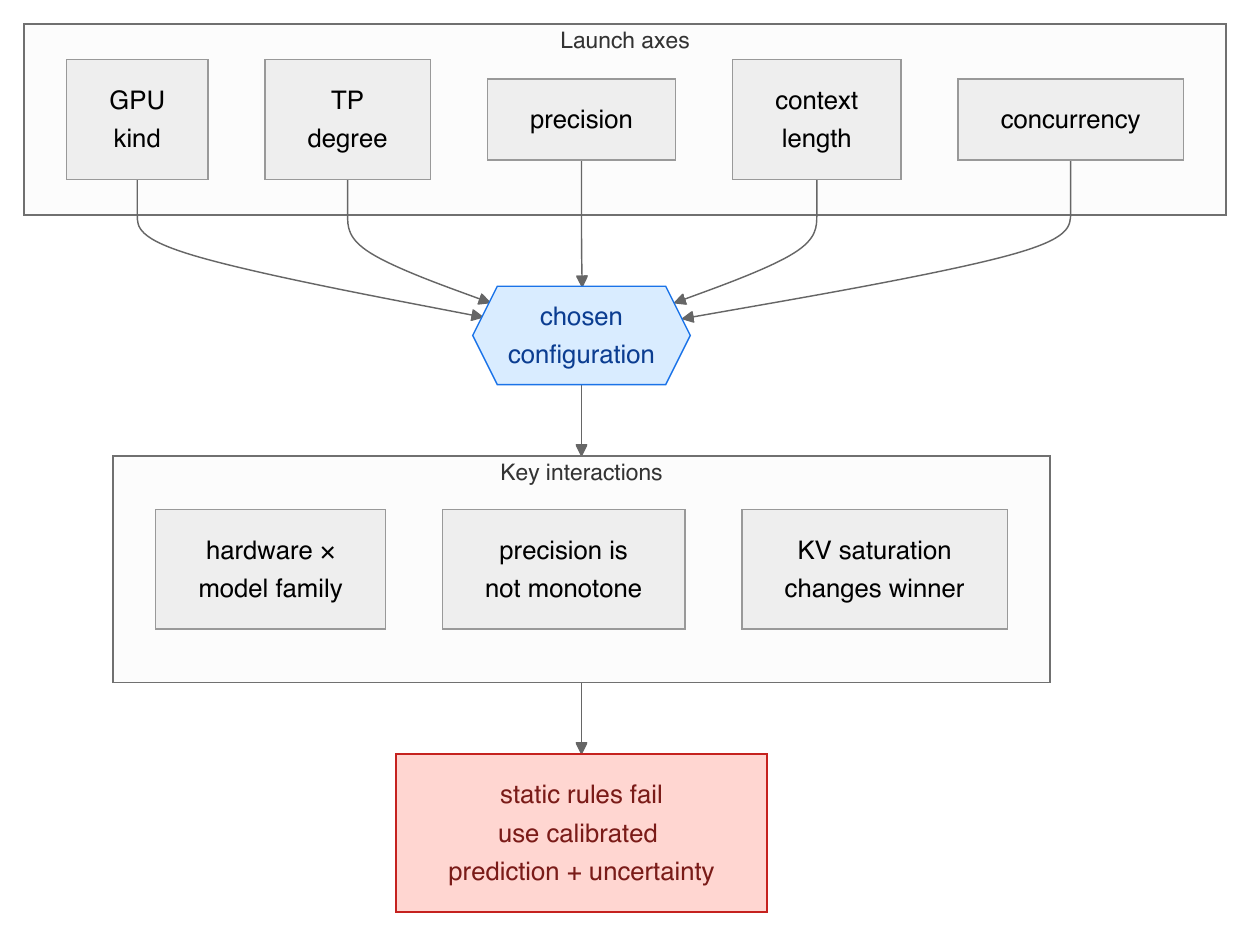}
\caption{The inference launch decision space (\S2.2, \S5.3). GPU kind,
tensor-parallel degree, precision, context, and concurrency define a
combinatorial space whose axes interact with the hardware ---
hardware$\times$family, non-monotone precision, and KV-saturation
crossovers --- in ways a static rule cannot capture.}
\label{fig:s4}
\end{figure}

Using these inputs, OmniPilot evaluates and selects from a combinatoric space of configuration arms defined by GPU kind $\times$ TP degree $\in\{1,2,4,8\}$ $\times$ precision $\in\{\text{bf16, FP8, AWQ, GPTQ}\}$ (Figure~\ref{fig:s4}). The output is a ranked recommendation accompanied by a calibrated cost interval per arm, or an \emph{abstention} when the request falls outside the system's measured support envelope. The optimization objective is expected economic utility: the predicted serving value net of node-hour cost, queue wait time, expected-failure restart cost, SLO violation probability, and a KV-saturation penalty.

\hypertarget{the-measurement-stale-problem}{%
\subsubsection{2.3 The measurement-stale problem}\label{the-measurement-stale-problem}}

Because the Data Center GPU Manager (DCGM) samples at a 60~s cadence, whole-job averages inadvertently capture low-activity container startup phases. This can severely understate actual serving-time behavior. For instance, restricting averages strictly to the active serving epoch increases measured SM activity from 21\% to 51\% and power draw from 132~W to 189~W. To prevent this skew, the advisor assigns explicit freshness windows to each metric (Appendix~C; Figure~\ref{fig:s3}) and actively down-weights stale measurements.

\begin{center}\rule{0.5\linewidth}{0.4pt}\end{center}

\hypertarget{method}{%
\subsection{3. Method}\label{method}}

\hypertarget{system-architecture}{%
\subsubsection{3.1 System Architecture}\label{system-architecture}}

The OmniPilot architecture (Figure~\ref{fig:arch}) is built around three core, evaluated components: the conformally calibrated cost model, the out-of-distribution abstention layer, and the decision-coupled placement advisor. (Note that two elements within the decision layer—the live resource query and the rule-based fallback triggered upon abstention—serve as operational scaffolding rather than evaluated contributions.)

The system's predictive intelligence is grounded in two distinct data stores. The personal store (\texttt{history.jsonl}, 1,667 typed rows) maintains the tightly controlled benchmark grid, while the cluster-wide store (\texttt{cluster\_history.jsonl}, 500K jobs) aggregates broad telemetry harvested across all users. To ensure strict data lineage and reproducibility, every logged record enforces a typed \texttt{kind}, a content hash for idempotent deduplication, a schema version, and a software-stack fingerprint. Appendix~C details the underlying collector schemas, gated model updates, recalibration, and benchmark-gate internals.

Two components of the decision layer are operational scaffolding rather
than evaluated contributions: the live resource query and the rule-based
fallback used after model abstention. The evaluated claims concern the
cost model, the abstention layer, and the decision-coupled placement
advisor.

Two stores support the system. The personal store
(\texttt{history.jsonl}, 1,667 typed rows) holds the controlled benchmark
grid. The cluster-wide store (\texttt{cluster\_history.jsonl}, 500K 
jobs) holds the all-users harvest. Every row carries a typed
\texttt{kind}, a content hash for idempotent deduplication, a schema
version, and a software-stack fingerprint. Appendix~C gives collector
schemas, gated model updates, recalibration, and benchmark-gate
internals.

\hypertarget{the-inference-cost-model}{%
\subsubsection{3.2 The Inference Cost
Model}\label{the-inference-cost-model}}

\hypertarget{design}{%
\paragraph{3.2.1 Design and feature engineering}\label{design}}

The cost model is a set of gradient-boosted regressors with quantile
loss (Figure~\ref{fig:model}, Appendix~A). For each target, it fits the
$\{0.1, 0.5, 0.9\}$ quantiles in log space because target values span
orders of magnitude and errors are multiplicative. The model predicts
eight targets. \textbf{Six drive decisions}: aggregate tokens/s, request
throughput, TTFT (p50), cold-start time, peak KV-cache usage, and average
power. \textbf{Two are diagnostic only}: SM- and tensor-core activity.
The 60 s DCGM cadence under-resolves these activity targets
($R^2$(log) 0.30--0.40), so they never enter the decision utility. A
target is fit only when it has at least fifteen labeled rows.
Appendix~B reports per-target feature exclusions, labeled-row counts,
and raw quantile-crossing diagnostics (Table~\ref{tab:targetcounts}).

The featurizer encodes roughly 28 features: log-transformed numerics
(model size, concurrency, context length, KV pressure, active MoE
parameters), linear numerics (TP degree, memory utilization, effective
bits), and one-hot encodings over nine model families, five GPU kinds,
and four quantization formats. A new axis must be represented
\emph{before} its data is collected. Removing the quantization feature
inflated out-of-fold FP8 error from 3.88\% to 7.36\% (+90\%) while
leaving bf16 unchanged (Figure~\ref{fig:quant}a; §5.4). The feature is
uninformative where precision does not vary and predictive where it does.

\hypertarget{calibration}{%
\paragraph{3.2.2 Calibration, abstention, and accuracy}\label{calibration}}

Prediction intervals use conformalized quantile regression (CQR)
{[}9{]}, {[}10{]}. For each target, OmniPilot applies a split-conformal
additive offset in log space, calibrated on held-out-cell residuals. This
grouped cross-conformal construction achieves 80\% coverage on unseen
workload cells, compared with 64--69\% under a random-fold offset. Beyond
the support envelope, conformal intervals fail (§4.3). OmniPilot therefore
uses support-distance abstention, rather than interval width alone, as
the out-of-distribution guardrail. Appendix~C.3 gives construction
details.

At fit time, the model stores its training envelope across seven axes. At
inference, a support check flags any out-of-envelope axis and downgrades
confidence to low, independent of interval width. Section~4.3 shows why
this support-distance signal is needed when interval width fails. The
measured envelope covers 3--9B models on A100/H100/H200 (51 cells, 96\%
top-1).

Five-fold cross-validation on 460 rows gives the per-target errors in
Table~\ref{tab:accuracy}. Figure~\ref{fig:accuracy} plots the same
breakdown.

\begin{longtable}[]{@{}llll@{}}
\caption{Cross-validated accuracy (460 rows).}\tabularnewline
\toprule
Target & MAPE & $R^2$ (log) & 80\% coverage\tabularnewline
\midrule
\endhead
\label{tab:accuracy}%
aggregate tokens/s & 6.2\% & 0.92 & 81\%\tabularnewline
request throughput & \textasciitilde{}7.6\% & 0.93 &
\textasciitilde{}81\%\tabularnewline
TTFT (p50) & \textasciitilde{}12\% & 0.92 & 80\%\tabularnewline
cold start & \textasciitilde{}11\% & 0.61 &
\textasciitilde{}79\%\tabularnewline
KV-cache usage & 25.3\% & \textasciitilde{}0.83 &
\textasciitilde{}66\%\tabularnewline
average power & \textasciitilde{}15\% & \textasciitilde{}0.46 &
---\tabularnewline
SM activity & \textasciitilde{}96\% & 0.34 & ---\tabularnewline
tensor activity & \textasciitilde{}100\% & 0.43 & ---\tabularnewline
\bottomrule
\end{longtable}

The main decision-driving latency and throughput targets have 6--12\%
MAPE with log-space $R^2$ of 0.92--0.93. KV-cache usage and average power
are less accurate but still decision-relevant. SM and tensor activity
have 96--100\% MAPE because the 60 s DCGM cadence under-resolves them;
they are diagnostic only.

\hypertarget{the-value-of-information-benchmark-gate}{%
\subsubsection{3.3 Decision Utility and the Launch-Success
Prior}\label{the-value-of-information-benchmark-gate}}

\hypertarget{economic-utility}{%
\paragraph{3.3.1 Economic utility}\label{economic-utility}}

OmniPilot ranks arms by utility. The utility equals predicted throughput
times runtime times a per-token value, minus compute cost, queue wait,
expected failure cost, SLO violation cost, and a KV-saturation penalty.
Compute cost is measured in capex-plus-power node-hours.

We calibrate value-per-token by revealed preference. The value
$2\times10^{-7}$ matches an expert operator's anchors
(7B$\to$A100, 32B$\to$2$\times$A100). At this value, joint placement
reaches 95\% top-1 accuracy with regret 0.003, compared with 0.020 under
a throughput-first setting.

A KV-saturation term applies a multiplicative throughput penalty above
an occupancy knee of 0.85. This converts KV preemptions into utility
cost. Adding the term corrected the Qwen2.5-32B/A100 precision decision
without adding data: precision backtest accuracy rose to 96\% top-1, and
the affected cell's regret dropped from 0.28.

\hypertarget{a-reliability-prior-from-cluster-data}{%
\paragraph{3.3.2 A launch-success prior from cluster
data}\label{a-reliability-prior-from-cluster-data}}

The failure term uses a per-configuration launch-success prior from the
cluster-wide harvest (§5.2). A beta-binomial estimator shrinks each
configuration toward the pooled rate, so sparsely sampled configurations
remain near the pool. Observed cluster success falls to about 38\% at
TP4, so multi-GPU configurations carry larger expected-failure cost than
under the flat 0.95 default. Appendix~B reports shrinkage sensitivity.

A regression-gated retraining loop promotes a candidate model only when
decision-target MAPE does not regress by more than 0.5 percentage
points. One observed cycle, from 447 to 460 rows, satisfied this gate
(Appendix~C.2).

\begin{center}\rule{0.5\linewidth}{0.4pt}\end{center}

\hypertarget{evaluation}{%
\subsection{4. Evaluation}\label{evaluation}}

\hypertarget{setup}{%
\subsubsection{4.1 Setup}\label{setup}}

The primary results use a frozen 460-row snapshot. It spans six model
families, three GPU kinds (A100-40GB, H100-80GB, H200-141GB), four
precisions, and TP degrees 1--8 (the full benchmark model list is in
Table~\ref{tab:modelzoo}, Appendix~B). Holdout cells vary context length (16k,
32k) and concurrency (1500, 2000), and are excluded from training. We
reran mechanism checks on the current 476-labeled-row dataset.

Backtests use leakage-free 5-fold GroupKFold grouped by workload cell:
base model $\times$ context $\times$ concurrency $\times$ prompt lengths.
A model is therefore never evaluated on a cell it trained on. Software is
pinned to vLLM 0.11.2. Each row also records a runtime fingerprint, so
the update loop can retire stale rows after runtime upgrades
(Appendix~C.2).

\hypertarget{precision-recommendation}{%
\subsubsection{4.2 Placement accuracy}\label{precision-recommendation}}

On 20 multi-precision cells, the advisor reaches 96\% top-1 accuracy and
0.001 regret. Always-bf16 has regret 0.144. This gain came from adding
the KV-aware utility term (§3.3.1), not from adding data; before the term,
the Qwen2.5-32B/A100 cell alone carried 0.28 regret. At the
cost-conscious operating point, precision accuracy is 88\% with 0.007
regret. In that setting, bf16 is more often sufficient because cost
dominates marginal throughput.

We backtested the full joint decision (GPU, TP, and precision) on 65
multi-arm workload cells from the frozen 460-row snapshot (§4.1). At the
cost-conscious calibrated value-per-token (§3.3.1), OmniPilot reaches 95\%
top-1 accuracy and regret 0.003. Bootstrap confidence intervals are
[89\%, 100\%] for top-1 accuracy and [0.000, 0.008] for regret. The
no-model baseline has regret 2.71, mainly because it over-provisions
Hopper GPUs when an A100 is sufficient. Per-band results
(Table~\ref{tab:sizeband}, Appendix~B) show that the 3--9B band has 51
cells and 96\% top-1 accuracy. Remaining misses are low-regret H100/H200
tie-breaks.

Table~\ref{tab:baselines} repeats the comparison on the current 476-row
dataset (66 cells; §4.1). OmniPilot reaches 92\% top-1 accuracy and regret
0.006, consistent with the frozen primary result. It has lower regret than the
size rule (1.97), H100/FP8 recipe (1.65), throughput-max policy (4.73),
and nearest-neighbor policy (0.72). Cheapest-fit is the closest baseline
with 91\% top-1 accuracy and regret 0.010. This baseline is competitive
because the calibrated operator preference is already cost-conscious,
but it cannot choose precision or anticipate KV-saturation cases.

\begin{table}[H]
\caption{Joint single-node placement: Comparing single-node placement performance between OmniPilot and model-free baselines. Evaluated via leakage-free GroupKFold across 66 workload cells, where lower regret indicates better performance.}\label{tab:baselines}
\centering\small
\begin{tabular}{@{}lrr@{}}
\toprule
Policy & Top-1 & Regret \\
\midrule
OmniPilot (advisor) & 92\% & 0.006 \\
Cheapest-fit & 91\% & 0.010 \\
Size-threshold TP rule & 80\% & 1.970 \\
H100/FP8 recipe & 42\% & 1.651 \\
Throughput-max (newest GPU) & 3\% & 4.728 \\
Nearest-neighbor cell & 70\% & 0.722 \\
OmniPilot (bf16-only) & 91\% & 0.010 \\
Mean-arm (random) & -- & 2.677 \\
\bottomrule
\end{tabular}
\end{table}

\hypertarget{out-of-distribution-abstention-1}{%
\subsubsection{4.3 Out-of-distribution behavior and calibration}\label{out-of-distribution-abstention-1}}

The holdout sweep used OLMo-2-13B as a new family and 13B size band. It
also used Qwen2.5-7B at 16k/32k context and 1500/2000 concurrency. All
five cells are outside the measured envelope. For each cell, the
canonical model predicted throughput before the run, and the run then
measured it (Table~\ref{tab:ood}).

The off-distribution predictions have 24--46\% error, compared with
roughly 6\% in envelope. The conformal intervals cover 0 of 5 points, so
interval width alone is not reliable beyond the envelope. The
support-distance signal flags all five cells as low-confidence. With only
five OOD cells, these numbers illustrate a failure regime rather than
estimate an OOD error rate. The supported claim is qualitative: beyond
the measured envelope, predictions fail and the abstention layer flags
the affected cells.

The confidence label also separates error regimes on in-distribution
traffic (Table~\ref{tab:confidence}, Appendix~B). A leave-one-family-out
test (Table~\ref{tab:generalization}, Appendix~B) shows that related
families predict to 11--19\% error, while novel architectures degrade to
29--126\% and are flagged.

\begin{longtable}[]{@{}lllllll@{}}
\caption{Out-of-distribution holdout: predict-before-run
(aggregate tokens/s).}\tabularnewline
\toprule
OOD cell & pred q50 & {[}q10, q90{]} & realized & error & covered? &
confidence\tabularnewline
\midrule
\endhead
\label{tab:ood}%
OLMo-13B (new family) & 17,026 & {[}14,560, 20,327{]} & 11,633 & 46\% &
no & low\tabularnewline
Qwen-7B ctx 16k & 16,583 & {[}8,669, 17,815{]} & 24,167 & 31\% & no &
low\tabularnewline
Qwen-7B ctx 32k & 16,538 & {[}6,545, 17,873{]} & 24,353 & 32\% & no &
low\tabularnewline
Qwen-7B conc 1500 & 23,174 & {[}8,774, 29,143{]} & 30,516 & 24\% & no &
low\tabularnewline
Qwen-7B conc 2000 & 23,663 & {[}8,774, 29,953{]} & 30,944 & 24\% & no &
low\tabularnewline
\bottomrule
\end{longtable}

Table~\ref{tab:recal} in Appendix~B evaluates the construction in §3.2.2.
Held-out-cell recalibration restores in-envelope coverage to 80--81\%,
compared with 64--69\% under random-fold calibration. The interval width
cost is 4--12\%. A disjoint-cell check in Appendix~C.3 gives the same
conclusion.

\hypertarget{ablations}{%
\subsubsection{4.4 Ablations and data efficiency}\label{ablations}}

Table~\ref{tab:ablation} in Appendix~B summarizes component ablations.
Each removal degrades the decision failure mode that the component
targets.

\begin{figure}[H]
\centering
\includegraphics[width=0.66\linewidth]{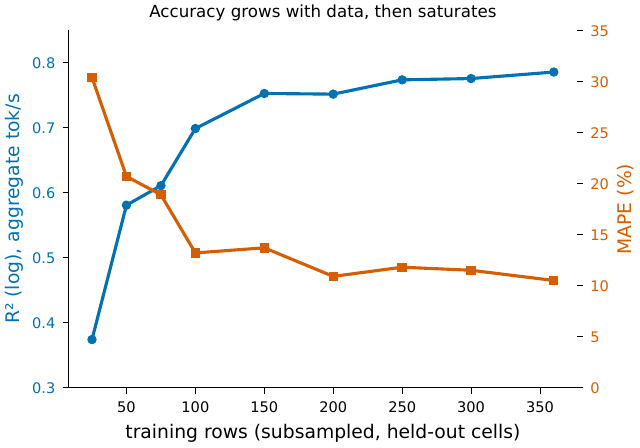}
\caption{Fixed-task learning curve (\S4.4): training rows are subsampled
and out-of-sample accuracy is measured on held-out workload cells (mean
over 12 seeds). Aggregate-throughput $R^2$ rises from 0.37 at 25 rows to
0.79 at 360 and median error falls from 30\% to 10\%, with gains
concentrated below 150 rows and saturating thereafter. The held-out-cell
ceiling sits below the five-fold $R^2$ of 0.92 because predicting an
unseen cell is harder than pooled cross-validation.}
\label{fig:s14}
\end{figure}

Accuracy tracks axis coverage more than raw row count. Adding 4-bit and
MoE data lifts $R^2$ from the bf16-only baseline to 0.923. In contrast,
a fixed-task learning curve shows diminishing returns above
$\sim$150 rows once the task is already covered
(Figures~\ref{fig:s10},~\ref{fig:s14}).

\hypertarget{threats-to-validity}{%
\subsubsection{4.5 Threats to validity}\label{threats-to-validity}}

The results come from a single cluster. Absolute thresholds, including
the 0.85 KV knee, freshness windows, and calibrated value-per-token, are
deployment-specific. The cluster-wide prior is observational and
confounded, so OmniPilot shrinks it toward the pooled rate and uses it only
as a launch-success and demand prior, not as placement ground truth.
Conformal intervals meet the in-envelope target after recalibration but
fail out-of-distribution (§4.3), so support-distance abstention is needed
outside the envelope. The update-loop evidence is a single promotion
cycle; it demonstrates the mechanism, not a longitudinal trend.

\begin{center}\rule{0.5\linewidth}{0.4pt}\end{center}

\hypertarget{discussion}{%
\subsection{5. Discussion}\label{discussion}}

\hypertarget{a-negative-result}{%
\subsubsection{5.1 Cluster telemetry: a negative result}\label{a-negative-result}}

The scheduler database sees all users, but it records utilization and
exit state rather than model, precision, tokens/s, or concurrency. We
harvest this all-user job history from Slurm's accounting database using
our HPC\_Tools collection scripts {[}28{]} and the \texttt{jobstats}
platform {[}29{]} (Appendix~C.1). Of
500K jobs, only 903 identify as vLLM-like. A cluster-wide throughput
advisor is therefore infeasible from infrastructure metadata alone. We
report this as an architectural finding and did not mine job scripts,
which would be privacy-sensitive.

\hypertarget{what-the-data-does-provide}{%
\subsubsection{5.2 What the cluster data provides}\label{what-the-data-does-provide}}

The harvest provides two usable signals (Figure~\ref{fig:cluster},
Appendix~B). First, demand is dominated by single-GPU jobs:
H100$\times$1 accounts for 43\% and A100$\times$1 for 38\%. The
benchmark grid covers every high-demand configuration. Second,
per-configuration launch success falls from 72\% at H100$\times$1 to
38\% at H100$\times$4. OmniPilot shrinks this signal toward the pooled rate
and uses it only in the utility's failure term (§3.3.2), not as a causal
hardware-reliability claim. Cross-validating against the cluster
population exposed the DCGM windowing artifact in §2.3. We did not blend
the two sources because the heavy mixed-job operating point differs from
light serving (Appendix~B).

\hypertarget{quantization-is-a-confounder-not-a-free-knob}{%
\subsubsection{5.3 Quantization introduces complex performance dynamics rather than monotonic gains}\label{quantization-is-a-confounder-not-a-free-knob}}

Numerical precisions do not follow a simple "fewer bits, faster" rule. While FP8 {[}8{]} increases throughput, AWQ {[}6{]} is slower on our kernel path, and GPTQ {[}7{]} fails entirely for Gemma-2-27b (Figure~\ref{fig:quant}). It is important to note that these performance characteristics reflect out-of-the-box vLLM configurations; while extensive tuning of internal engine parameters might alter these dynamics, we model the baseline operational reality. To capture these nuances, OmniPilot pairs a monotonic effective-bits feature for general compression trends with per-format one-hot features to isolate off-trend behaviors like AWQ (evaluated in §4.2). 

KV-cache pressure introduces a second complex interaction driven by cache saturation rather than raw model size. For example, running Qwen2.5-32B on an A100 in bf16 saturates its KV cache (1.00 occupancy), triggering frequent preemptions that spike TTFT to $\sim$16~s. Under identical conditions, Mistral-Small-24B maintains sufficient headroom (0.53 occupancy), leaving bf16 as the preferred choice. This non-linear crossover behavior directly motivates the explicit KV-saturation penalty within our decision utility function (§3.3.1).

\hypertarget{feature-before-data-as-a-general-method}{%
\subsubsection{5.4 Feature-before-data as a general
method}\label{feature-before-data-as-a-general-method}}

We draw three lessons from the advisor. First, represent a new axis in
the featurizer \emph{before} collecting data on it. The quantization
ablation in §3.2.1 shows how this test separates unused features from
features that become predictive once the axis varies. Second, some
improvements belong in the utility or measurement pipeline rather than
the model weights. The KV-saturation term and the DCGM serving-window
fix are examples. Third, scheduler metadata can support demand and
launch-success priors but cannot recover throughput without workload
metadata. A cluster-wide throughput advisor therefore requires
instrumentation at the serving layer.

The benchmark gate (Appendix~C.4) returns zero value of information at
the cost-conscious operating point. This is a structural limitation of
the current conditioning rule, which shrinks predictive variance without
updating the posterior mean. A useful gate needs posterior mean updates
or an equivalent refitting approximation (§5.5).

\hypertarget{scope-and-future-work}{%
\subsubsection{5.5 Scope and future work}\label{scope-and-future-work}}

The primary direction for future research is expanding coverage across broader ML workloads. While OmniPilot can reuse its existing telemetry substrate, extending the system to support training and fault recovery will require developing new targets, utilities, and labels. On the modeling side, the most immediate next step is implementing a benchmark gate that updates posterior means (§5.4). Furthermore, because our current evaluation is constrained to a single-cluster setting, future deployments will require revalidating thresholds across diverse cluster environments. Finally, conducting a longitudinal update-and-drift study over multiple promotion cycles will be essential for assessing the system's behavioral stability over time.

While the current model successfully leverages cluster telemetry and Slurm job parameters for vLLM inference workloads, it does not currently account for internal engine configurations. Future work will explore how specific vLLM runtime parameters influence inference performance, aiming to integrate these variables into the predictive model to further optimize serving decisions.

\begin{center}\rule{0.5\linewidth}{0.4pt}\end{center}

\hypertarget{related-work}{%
\subsection{6. Related Work}\label{related-work}}

\textbf{Serving systems.} vLLM {[}1{]}, Orca {[}2{]}, Sarathi-Serve
{[}3{]}, SGLang {[}4{]}, and AlpaServe {[}5{]} optimize execution within
a chosen configuration. Recent work optimizes already-chosen deployments
through prefill/decode disaggregation {[}21{]}, {[}22{]},
KV-cache-centric disaggregation {[}25{]}, cross-instance migration
{[}23{]}, heterogeneous routing {[}24{]}, and paging-free KV management
{[}26{]}. These systems take the model, hardware, and parallelism as
given. OmniPilot is upstream: it chooses GPU, TP degree, and precision
\emph{before} submission and abstains outside its measured envelope.
Helix {[}24{]} is closest in spirit, but it routes over a fixed fleet and
does not model quantization or calibrated uncertainty.

\textbf{Quantization and uncertainty.} FP8 {[}8{]}, GPTQ {[}7{]}, and
AWQ {[}6{]} characterize accuracy retention. OmniPilot instead estimates
when each precision is beneficial on a given hardware path. The model
captures AWQ's off-trend slowness and hard incompatibilities as measured
failure modes. OmniPilot's intervals use CQR {[}9{]}, {[}10{]}, paired
with abstention because coverage degrades out-of-distribution. Its
benchmark gate applies information-value theory {[}11{]} to decide
whether a pre-launch probe is worth its node-hours.

\textbf{Performance modeling and scheduling.} Habitat {[}13{]},
Roofline {[}14{]}, and MLPerf {[}15{]} predict or bound
training/inference performance analytically. OmniPilot fits an empirical,
decision-coupled model instead. The gating discipline follows
production-ML validation {[}16{]}, {[}17{]} and continual-learning
regression avoidance {[}18{]}. Cluster schedulers such as Gandiva
{[}19{]} and Tiresias {[}20{]} are complementary: they act at submission,
while OmniPilot advises before submission.

\begin{center}\rule{0.5\linewidth}{0.4pt}\end{center}

\hypertarget{conclusion}{%
\subsection{7. Conclusion}\label{conclusion}}

OmniPilot optimizes inference workloads within shared, multi-tenant, heterogeneous GPU clusters by transforming hardware configuration selection into a calibrated economic decision. Recognizing that cluster operators have fluctuating priorities—such as weighting power efficiency over token throughput—the system features a flexible utility function that allows optimization goals to be dynamically adjusted along any axis. Acting as a macro-level planner, OmniPilot complements specialized, hardware-level inference engines like vLLM, SGLang, and TensorRT. By combining a conformally calibrated quantile cost model, an out-of-distribution (OOD) abstention layer, a cluster-wide launch-success prior, and a regression-gated update loop, the system reduces placement regret by more than an order of magnitude while its safety shield reliably flags 100

The development of this system also yields three fundamental methodological lessons for cluster-scale machine learning infrastructure. First, new operational axes must be explicitly integrated into the featurizer before data collection begins rather than retroactively. Second, many prediction errors are more effectively resolved through refined utility design or measurement precision rather than simply expanding model capacity. Finally, while standard scheduler metadata (like Slurm logs) successfully provides demand and launch-success priors, it cannot support a cluster-wide throughput advisor because it only records resource allocation—how a GPU was occupied—rather than the specific software workload running on it.

Looking ahead, the primary future direction is expanding OmniPilot's coverage to broader ML workloads. While the current framework is limited to single-node inference serving, extending the system to handle training and fault-recovery workloads will reuse the existing telemetry substrate while introducing new target metrics and utility functions. Additionally, future research will focus on expanding the planner to support multi-node serving architectures and conducting longitudinal update studies to evaluate how the system manages data drift over extended operational periods.

\begin{center}\rule{0.5\linewidth}{0.4pt}\end{center}

\section*{Acknowledgments}
This work was made possible in part by a gift from the Chan Zuckerberg Initiative Foundation to establish the Kempner Institute for the Study of Natural and Artificial Intelligence at Harvard University. We also thank the Kempner Institute Research Engineering team and the Faculty of Arts and Sciences Research Computing staff for their valuable discussions and support.

\hypertarget{declarations}{%
\subsection{Declarations}\label{declarations}}

\begin{itemize}
\tightlist
\item
  \textbf{Data and code availability.} The benchmark dataset, cost
  model, and advisor code are in the project repository; the
  cluster-wide harvest stores users as salted hashes and is not
  redistributable.
\item
  \textbf{Conflict of interest.} The authors declare no competing
  interests.
\item
  \textbf{Ethics.} The cluster-wide harvest uses only scheduler
  metadata and DCGM metrics, anonymizes users via salted hashing, and does not read job
  scripts, logs, or payloads.
\item
  \textbf{AI-usage disclosure.} Draft prepared with AI assistance for
  structuring and prose; all quantitative claims derive from the
  authors' measured experiments and all citations were verified against
  primary sources. A venue-specific statement will be generated at
  submission.
\item
  \textbf{Limitations.} See §4.5 and §5.4--5.5.
\end{itemize}

\hypertarget{references}{%
\subsection{References}\label{references}}

{[}1{]} W. Kwon, Z. Li, S. Zhuang, Y. Sheng, L. Zheng, C. H. Yu, J. E.
Gonzalez, H. Zhang, and I. Stoica, ``Efficient Memory Management for
Large Language Model Serving with PagedAttention,'' in \emph{Proc. 29th
ACM Symp. Operating Systems Principles (SOSP)}, 2023.
doi:10.1145/3600006.3613165. arXiv:2309.06180.

{[}2{]} G.-I. Yu, J. S. Jeong, G.-W. Kim, S. Kim, and B.-G. Chun,
``Orca: A Distributed Serving System for Transformer-Based Generative
Models,'' in \emph{Proc. 16th USENIX Symp. Operating Systems Design and
Implementation (OSDI)}, 2022.

{[}3{]} A. Agrawal, N. Kedia, A. Panwar, J. Mohan, N. Kwatra, B. S.
Gulavani, A. Tumanov, and R. Ramjee, ``Taming Throughput-Latency
Tradeoff in LLM Inference with Sarathi-Serve,'' in \emph{Proc. 18th
USENIX Symp. Operating Systems Design and Implementation (OSDI)}, 2024.
arXiv:2403.02310.

{[}4{]} L. Zheng, L. Yin, Z. Xie, C. Sun, J. Huang, C. H. Yu, S. Cao, C.
Kozyrakis, I. Stoica, J. E. Gonzalez, C. Barrett, and Y. Sheng,
``SGLang: Efficient Execution of Structured Language Model Programs,''
in \emph{Advances in Neural Information Processing Systems (NeurIPS)},
2024. arXiv:2312.07104.

{[}5{]} Z. Li, L. Zheng, Y. Zhong, V. Liu, Y. Sheng, X. Jin, Y. Huang,
Z. Chen, H. Zhang, J. E. Gonzalez, and I. Stoica, ``AlpaServe:
Statistical Multiplexing with Model Parallelism for Deep Learning
Serving,'' in \emph{Proc. 17th USENIX Symp. Operating Systems Design and
Implementation (OSDI)}, 2023. arXiv:2302.11665.

{[}6{]} J. Lin, J. Tang, H. Tang, S. Yang, W.-M. Chen, W.-C. Wang, G.
Xiao, X. Dang, C. Gan, and S. Han, ``AWQ: Activation-aware Weight
Quantization for LLM Compression and Acceleration,'' in \emph{Proc.
Machine Learning and Systems (MLSys)}, 2024. arXiv:2306.00978.

{[}7{]} E. Frantar, S. Ashkboos, T. Hoefler, and D. Alistarh, ``GPTQ:
Accurate Post-Training Quantization for Generative Pre-trained
Transformers,'' in \emph{Proc. Int. Conf. Learning Representations
(ICLR)}, 2023. arXiv:2210.17323.

{[}8{]} P. Micikevicius, D. Stosic, N. Burgess, M. Cornea, P. Dubey, R.
Grisenthwaite, S. Ha, A. Heinecke, P. Judd, J. Kamalu et al., ``FP8
Formats for Deep Learning,'' arXiv:2209.05433, 2022.

{[}9{]} Y. Romano, E. Patterson, and E. J. Candès, ``Conformalized
Quantile Regression,'' in \emph{Advances in Neural Information
Processing Systems (NeurIPS)}, 2019, pp.~3538--3548. arXiv:1905.03222.

{[}10{]} A. N. Angelopoulos and S. Bates, ``A Gentle Introduction to
Conformal Prediction and Distribution-Free Uncertainty Quantification,''
arXiv:2107.07511, 2021.

{[}11{]} R. A. Howard, ``Information Value Theory,'' \emph{IEEE Trans.
Systems Science and Cybernetics}, vol.~2, no. 1, pp.~22--26, 1966.
doi:10.1109/TSSC.1966.300074.

{[}12{]} F. M. Polo, L. Weber, L. Choshen, Y. Sun, G. Xu, and M.
Yurochkin, ``tinyBenchmarks: Evaluating LLMs with Fewer Examples,'' in
\emph{Proc. Int. Conf. Machine Learning (ICML)}, 2024. arXiv:2402.14992.

{[}13{]} G. X. Yu, Y. Gao, P. Golikov, and G. Pekhimenko, ``Habitat: A
Runtime-Based Computational Performance Predictor for Deep Neural
Network Training,'' in \emph{Proc. USENIX Annual Technical Conf. (ATC)},
2021. arXiv:2102.00527.

{[}14{]} S. Williams, A. Waterman, and D. Patterson, ``Roofline: An
Insightful Visual Performance Model for Multicore Architectures,''
\emph{Communications of the ACM}, vol.~52, no. 4, pp.~65--76, 2009.
doi:10.1145/1498765.1498785.

{[}15{]} V. J. Reddi, C. Cheng, D. Kanter, P. Mattson, G. Schmuelling,
C.-J. Wu et al., ``MLPerf Inference Benchmark,'' in \emph{Proc. ACM/IEEE
47th Int. Symp. Computer Architecture (ISCA)}, 2020.
doi:10.1109/ISCA45697.2020.00045. arXiv:1911.02549.

{[}16{]} D. Baylor, E. Breck, H.-T. Cheng, N. Fiedel, C. Y. Foo, Z.
Haque, S. Haykal, M. Ispir, V. Jain, L. Koc et al., ``TFX: A
TensorFlow-Based Production-Scale Machine Learning Platform,'' in
\emph{Proc. 23rd ACM SIGKDD Int. Conf. Knowledge Discovery and Data
Mining (KDD)}, 2017. doi:10.1145/3097983.3098021.

{[}17{]} E. Breck, N. Polyzotis, S. Roy, S. E. Whang, and M. Zinkevich,
``Data Validation for Machine Learning,'' in \emph{Proc. Machine
Learning and Systems (MLSys)}, 2019.

{[}18{]} M. De Lange, R. Aljundi, M. Masana, S. Parisot, X. Jia, A.
Leonardis, G. Slabaugh, and T. Tuytelaars, ``A Continual Learning
Survey: Defying Forgetting in Classification Tasks,'' \emph{IEEE Trans.
Pattern Analysis and Machine Intelligence}, 2021. arXiv:1909.08383.

{[}19{]} W. Xiao, R. Bhardwaj, R. Ramjee, M. Sivathanu, N. Kwatra, Z.
Han, P. Patel, X. Peng, H. Zhao, Q. Zhang, F. Yang, and L. Zhou,
``Gandiva: Introspective Cluster Scheduling for Deep Learning,'' in
\emph{Proc. 13th USENIX Symp. Operating Systems Design and
Implementation (OSDI)}, 2018.

{[}20{]} J. Gu, M. Chowdhury, K. G. Shin, Y. Zhu, M. Jeon, J. Qian, H.
Liu, and C. Guo, ``Tiresias: A GPU Cluster Manager for Distributed Deep
Learning,'' in \emph{Proc. 16th USENIX Symp. Networked Systems Design
and Implementation (NSDI)}, 2019.

{[}21{]} Y. Zhong, S. Liu, J. Chen, J. Hu, Y. Zhu, X. Liu, X. Jin, and H.
Zhang, ``DistServe: Disaggregating Prefill and Decoding for
Goodput-optimized Large Language Model Serving,'' in \emph{Proc. 18th
USENIX Symp. Operating Systems Design and Implementation (OSDI)}, 2024.
arXiv:2401.09670.

{[}22{]} P. Patel, E. Choukse, C. Zhang, A. Shah, Í. Goiri, S. Maleki,
and R. Bianchini, ``Splitwise: Efficient Generative LLM Inference Using
Phase Splitting,'' in \emph{Proc. 51st ACM/IEEE Int. Symp. Computer
Architecture (ISCA)}, 2024. arXiv:2311.18677.

{[}23{]} B. Sun, Z. Huang, H. Zhao, W. Xiao, X. Zhang, Y. Li, and W. Lin,
``Llumnix: Dynamic Scheduling for Large Language Model Serving,'' in
\emph{Proc. 18th USENIX Symp. Operating Systems Design and
Implementation (OSDI)}, 2024. arXiv:2406.03243.

{[}24{]} Y. Mei, Y. Zhuang, X. Miao, J. Yang, Z. Jia, and R. Vinayak,
``Helix: Serving Large Language Models over Heterogeneous GPUs and
Network via Max-Flow,'' in \emph{Proc. 30th ACM Int. Conf. Architectural
Support for Programming Languages and Operating Systems (ASPLOS)}, 2025.
doi:10.1145/3669940.3707215. arXiv:2406.01566.

{[}25{]} R. Qin, Z. Li, W. He, M. Zhang, Y. Wu, W. Zheng, and X. Xu,
``Mooncake: A KVCache-centric Disaggregated Architecture for LLM
Serving,'' in \emph{Proc. 23rd USENIX Conf. File and Storage
Technologies (FAST)}, 2025. arXiv:2407.00079.

{[}26{]} R. Prabhu, A. Nayak, J. Mohan, R. Ramjee, and A. Panwar,
``vAttention: Dynamic Memory Management for Serving LLMs without
PagedAttention,'' in \emph{Proc. 30th ACM Int. Conf. Architectural
Support for Programming Languages and Operating Systems (ASPLOS)}, 2025.
doi:10.1145/3669940.3707256. arXiv:2405.04437.

{[}27{]} Kempner Institute for the Study of Natural and Artificial
Intelligence, ``The Kempner AI Cluster,'' Harvard University.
\url{https://kempnerinstitute.harvard.edu/kempner-ai-cluster/}.

{[}28{]} D. Balamurugan, ``HPC\_Tools: Utilities for HPC Cluster
Telemetry and Job-Data Collection,'' GitHub repository.
\url{https://github.com/dmbala/HPC_Tools}.

{[}29{]} Princeton University Research Computing, ``Jobstats: A Slurm
Job Statistics and Monitoring Platform,'' GitHub repository.
\url{https://github.com/PrincetonUniversity/jobstats}.

\begin{center}\rule{0.5\linewidth}{0.4pt}\end{center}

\setcounter{figure}{0}
\setcounter{table}{0}
\renewcommand{\thefigure}{A\arabic{figure}}
\renewcommand{\thetable}{A\arabic{table}}

\hypertarget{supplementary-figures}{%
\subsection{Appendix A: Supplementary
Figures}\label{supplementary-figures}}

The following figures supplement the main text; each is referenced from
its corresponding section, and all numbers trace to the same measured
results as the body.

\begin{figure}[H]
\centering
\includegraphics[width=0.45\linewidth]{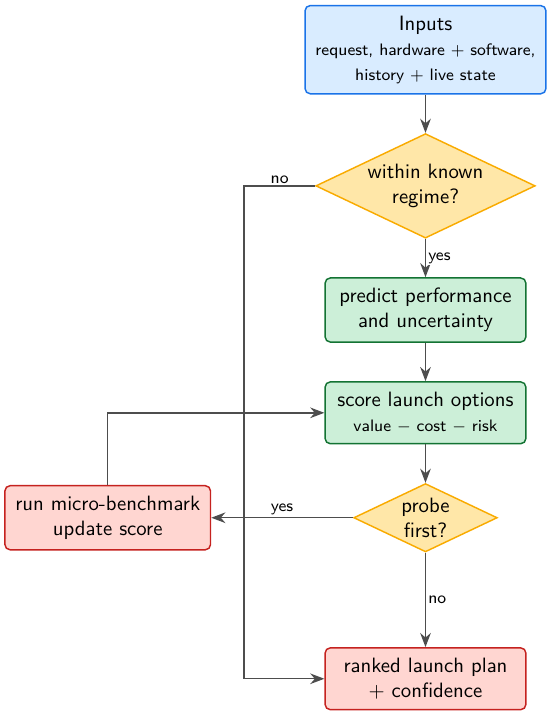}
\caption{\textbf{From features to a ranked launch (\S3.2.1).} Inputs (the request,
hardware and software, and history plus live cluster state) are first
checked against the model's support envelope. An in-regime request is
scored by predicted performance and uncertainty under a
value-minus-cost-minus-risk utility; the benchmark gate then decides
whether a cheap pre-launch probe would change the ranking, looping back
to re-score if so. An out-of-regime request bypasses prediction and is
flagged. The output is a ranked launch plan with a confidence label
(§3.2, §3.3; the gate is analyzed in §5.4).}
\label{fig:model}
\end{figure}

\begin{figure}[H]
\centering
\includegraphics[width=\linewidth]{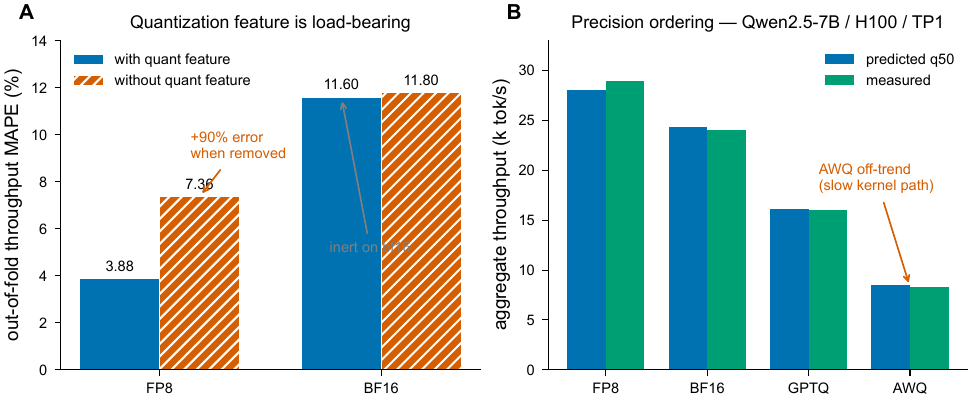}
\caption{Quantization introduces complex performance dynamics rather than monotonic gains. (a) Removing the quantization feature nearly doubles out-of-fold FP8 throughput error (3.88\% $\to$ 7.36\%) while leaving bf16 unchanged. This confirms the feature is highly predictive wherever precision varies, while remaining safely uninformative where it does not. (b) On Qwen2.5-7B/H100/TP1, the model recovers the full, non-monotone precision ordering. Crucially, it captures AWQ's off-trend slowness, an anomaly that a strictly monotonic bit-width feature would miss.}
\label{fig:quant}
\end{figure}

\begin{figure}[H]
\centering
\includegraphics[width=0.32\linewidth]{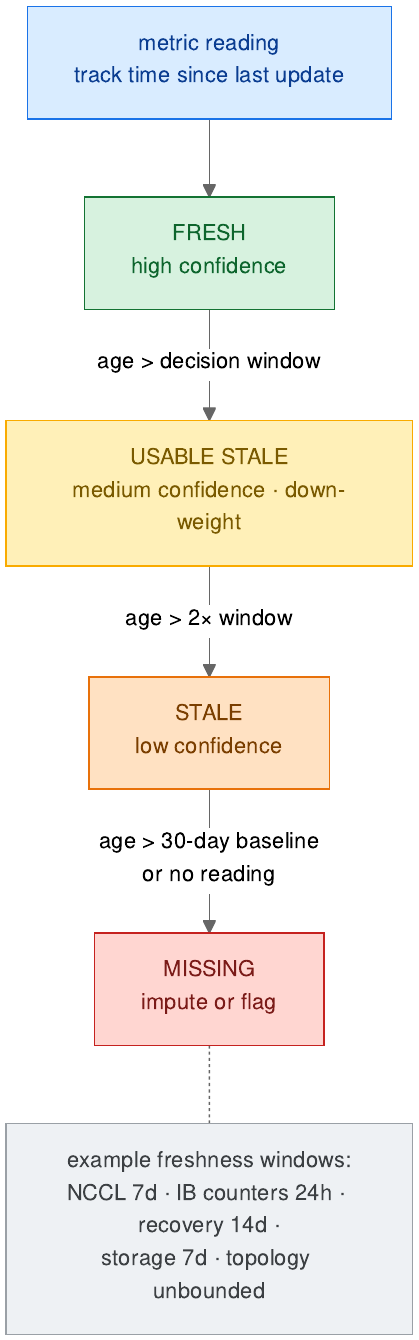}
\caption{Per-metric freshness state machine (\S2.3). Each
telemetry reading is labeled fresh, usable-stale, stale, or missing by
its age against a metric-specific decision window (e.g.\ NCCL 7d,
InfiniBand counters 24h, recovery 14d), so a stale measurement is
down-weighted rather than trusted at face value.}
\label{fig:s3}
\end{figure}

\begin{figure}[H]
\centering
\includegraphics[width=0.66\linewidth]{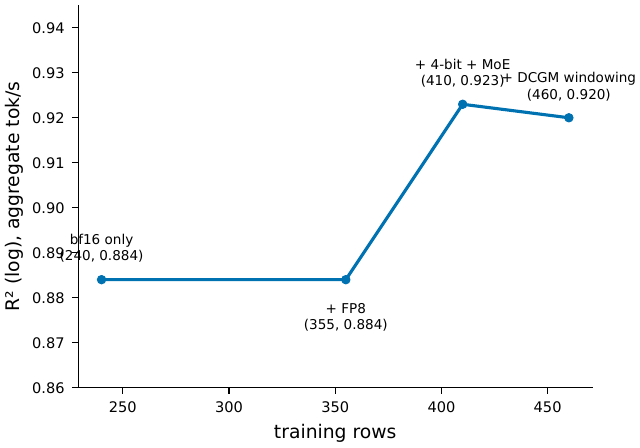}
\caption{Data efficiency, axis-addition view (\S4.4): aggregate-throughput
$R^2$ tracks coverage of the feature axes, not raw row count --- adding
FP8 to a bf16-only model leaves $R^2$ flat (precision already
featurized), while adding 4-bit and MoE coverage lifts it. Figure~\ref{fig:s14}
gives the complementary fixed-task learning curve.}
\label{fig:s10}
\end{figure}

\begin{figure}[H]
\centering
\includegraphics[width=\linewidth]{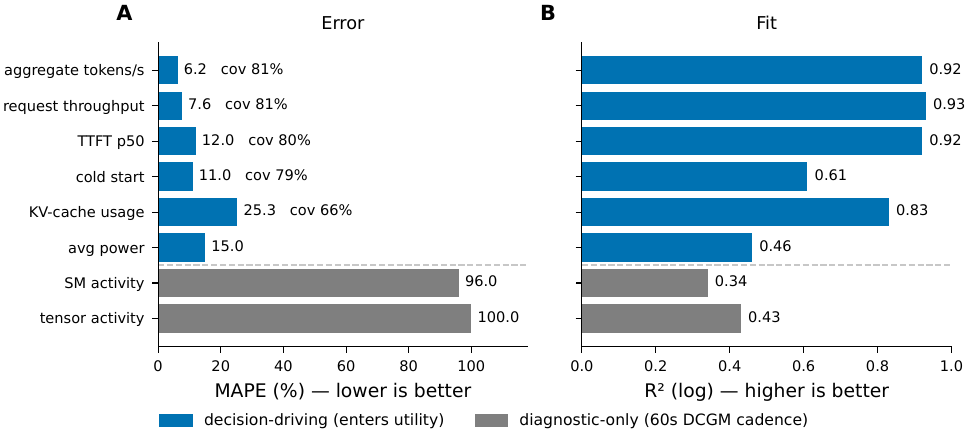}
\caption{Cross-validated cost-model accuracy across the eight targets
(5-fold, 460 rows; the per-target breakdown of Table~\ref{tab:accuracy}, \S3.2.2). The six
decision-driving targets (blue) are predicted to 6--25\% MAPE with high
log-space $R^2$ and near-nominal 80\% interval coverage; the two
diagnostic-only targets (grey) are poorly resolved by the 60\,s DCGM
cadence and are excluded from the decision utility.}
\label{fig:accuracy}
\end{figure}

\begin{figure}[H]
\centering
\includegraphics[width=\linewidth]{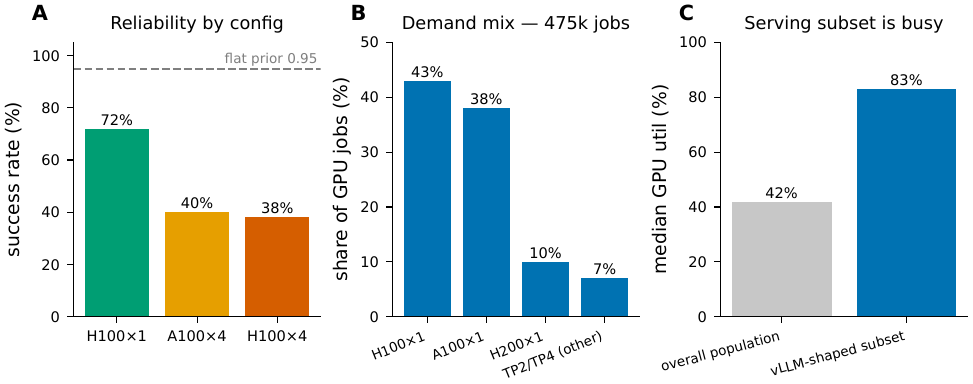}
\caption{What the 475,398-job cluster harvest provides (\S5.2). (a)
Per-configuration success rate falls from 72\% at H100$\times$1 to 38\%
at H100$\times$4, against the advisor's flat 0.95 prior. (b) Demand is
single-GPU-dominated. (c) The serving-shaped subset is busy, with median
GPU utilization 83\% versus 42\% overall.}
\label{fig:cluster}
\end{figure}

\begin{center}\rule{0.5\linewidth}{0.4pt}\end{center}

\setcounter{figure}{0}
\setcounter{table}{0}
\renewcommand{\thefigure}{B\arabic{figure}}
\renewcommand{\thetable}{B\arabic{table}}

\hypertarget{supplementary-tables}{%
\subsection{Appendix B: Supplementary
Tables}\label{supplementary-tables}}

These tables report the evaluation on the \emph{current} snapshot
(476 labeled inference rows parsed from the 561-run grid). The frozen
460-row backtest in §4 is unchanged.

\begin{table}[H]
\caption{
Held-out-cell conformal recalibration: Applying an offset calibrated on held-out residuals restores in-envelope coverage to the 80\% target, at the cost of 4--12\% wider prediction intervals.}\label{tab:recal}
\centering\small
\begin{tabular}{@{}lrrr@{}}
\toprule
Target & random-fold cov. & held-out-cell cov. & width $\times$ \\
\midrule
aggregate tokens/s & 64\% & \textbf{80\%} & 1.12 \\
TTFT (p50) & 69\% & \textbf{80\%} & 1.08 \\
KV-cache usage & 69\% & \textbf{81\%} & 1.04 \\
\bottomrule
\end{tabular}
\end{table}

\begin{longtable}[]{@{}ll@{}}
\caption{Component ablations.}\tabularnewline
\toprule
\begin{minipage}[b]{0.47\columnwidth}\raggedright
Remove\strut
\end{minipage} & \begin{minipage}[b]{0.47\columnwidth}\raggedright
Effect\strut
\end{minipage}\tabularnewline
\midrule
\endhead
\label{tab:ablation}%
\begin{minipage}[t]{0.47\columnwidth}\raggedright
quantization feature\strut
\end{minipage} & \begin{minipage}[t]{0.47\columnwidth}\raggedright
FP8 throughput error 3.88\% -\textgreater{} 7.36\% (+90\%); bf16
unchanged\strut
\end{minipage}\tabularnewline
\begin{minipage}[t]{0.47\columnwidth}\raggedright
KV-saturation utility term\strut
\end{minipage} & \begin{minipage}[t]{0.47\columnwidth}\raggedright
Qwen2.5-32B/A100 miss returns (cell regret 0.28)\strut
\end{minipage}\tabularnewline
\begin{minipage}[t]{0.47\columnwidth}\raggedright
per-config success prior\strut
\end{minipage} & \begin{minipage}[t]{0.47\columnwidth}\raggedright
over-recommends multi-GPU configs whose real success is
\textasciitilde{}38\%\strut
\end{minipage}\tabularnewline
\begin{minipage}[t]{0.47\columnwidth}\raggedright
model-family encodings\strut
\end{minipage} & \begin{minipage}[t]{0.47\columnwidth}\raggedright
gemma-2-27b throughput prediction degrades (family is top feature for
KV)\strut
\end{minipage}\tabularnewline
\begin{minipage}[t]{0.47\columnwidth}\raggedright
promotion gate\strut
\end{minipage} & \begin{minipage}[t]{0.47\columnwidth}\raggedright
provenance counterfactual ships a higher-error model when data grows
noisier\strut
\end{minipage}\tabularnewline
\bottomrule
\end{longtable}

\begin{table}[H]
\caption{Placement accuracy by model-size band (65 cells, calibrated operating point, \S4.2).}\label{tab:sizeband}
\centering\small
\begin{tabular}{@{}lllll@{}}
\toprule
Size band & Cells & Top-1 & Regret & Supported \\
\midrule
$\le$2B & 3 & 100\% & 0.000 & thin \\
\textbf{3--9B} & \textbf{51} & \textbf{96\%} & \textbf{0.003} & \textbf{yes} \\
10--23B & 2 & 100\% & 0.000 & thin \\
$\ge$24B & 9 & 89\% & 0.008 & thin \\
\bottomrule
\end{tabular}
\end{table}

\begin{table}[H]
\caption{Confidence-label calibration on the leakage-free
GroupKFold folds (current dataset, 261 out-of-fold predictions). The
labels separate regimes of materially different error; \emph{low} is the
abstention band. Abstention rate 3\% on in-distribution traffic.}\label{tab:confidence}
\centering\small
\begin{tabular}{@{}lrrrr@{}}
\toprule
Confidence & $n$ & MAPE & 80\% coverage & q10--q90 width \\
\midrule
High & 201 & 18\% & 69\% & 44\% \\
Medium & 52 & 146\% & 69\% & 143\% \\
Low (abstain) & 8 & 39\% & 62\% & 125\% \\
\bottomrule
\end{tabular}
\end{table}

\begin{table}[H]
\caption{Per-target labeled-row counts and raw quantile-crossing
rate (q10$>$q50 or q50$>$q90, before the monotone rearrangement that
prediction applies). Per-target feature exclusions are fixed a priori,
not selected inside folds.}\label{tab:targetcounts}
\centering\small
\begin{tabular}{@{}lrr@{}}
\toprule
Target & labeled rows & crossing rate \\
\midrule
aggregate tok/s & 476 & 9.7\% \\
request throughput & 476 & 8.4\% \\
TTFT (p50) & 476 & 1.9\% \\
cold start & 472 & 1.1\% \\
KV-cache usage & 329 & 10.1\% \\
average power & 476 & 0.4\% \\
SM activity & 476 & 6.1\% \\
tensor activity & 476 & 10.1\% \\
\bottomrule
\end{tabular}
\end{table}

\begin{table}[H]
\caption{Benchmark model zoo summarized by model family: number of
distinct models, parameter-size range, precision formats exercised, and
total benchmark runs. Coverage is uneven by design --- a few anchor
models (notably Qwen2.5-7B, 167 runs) carry most rows. Most models were
benchmarked on all three GPU types (A100, H100, H200); small
($\le$1.5B), sparse-MoE, and held-out probe models ran on a Hopper
subset. Runs sum to the 476 labeled rows.}\label{tab:modelzoo}
\centering\small
\begin{tabular}{@{}lrllr@{}}
\toprule
Family & Models & Size (B) & Precisions & Runs \\
\midrule
qwen2 & 6 & 0.5--32 & awq/bf16/fp8/gptq & 236 \\
llama3 & 2 & 8 & awq/bf16/fp8/gptq & 89 \\
deepseek\_distill & 1 & 7 & bf16/fp8 & 67 \\
mistral & 4 & 7--46.7 & awq/bf16/gptq & 35 \\
phi & 4 & 3.8--41.9 & bf16 & 23 \\
gemma & 3 & 2--27 & bf16 & 19 \\
other (pythia, olmo) & 2 & 12--13 & bf16 & 7 \\
\midrule
Total & 22 & 0.5--46.7 & awq/bf16/fp8/gptq & 476 \\
\bottomrule
\end{tabular}
\end{table}

\begin{table}[H]
\caption{Generalization to unseen models (leave-one-out,
aggregate tok/s). Held out and predicted from the rest; compare with
$\approx$6\% in-distribution. Families/models close to a seen
architecture interpolate; novel architectures are out-of-distribution and
trip the support-distance abstention (\S4.3).}\label{tab:generalization}
\centering\small
\begin{tabular}{@{}lrr@{}}
\toprule
Held-out family & $n$ & MAPE \\
\midrule
deepseek\_distill & 67 & 11\% \\
llama3 & 89 & 14\% \\
mistral & 35 & 16\% \\
qwen2 & 236 & 19\% \\
other (pythia, olmo) & 7 & 29\% \\
gemma & 19 & 56\% \\
phi & 23 & 126\% \\
\midrule
\multicolumn{3}{@{}l}{\emph{Specific held-out model}} \\
Qwen2-7B-Instruct & 4 & 2\% \\
Llama-3.1-8B-Instruct & 11 & 2\% \\
gemma-2-2b-it & 6 & 5\% \\
Phi-3-mini-4k-instruct-3.8B & 6 & 63\% \\
gemma-2-9b-it & 7 & 66\% \\
gemma-2-27b-it & 6 & 67\% \\
\bottomrule
\end{tabular}
\end{table}

\begin{center}\rule{0.5\linewidth}{0.4pt}\end{center}

\setcounter{figure}{0}
\setcounter{table}{0}
\renewcommand{\thefigure}{C\arabic{figure}}
\renewcommand{\thetable}{C\arabic{table}}

\hypertarget{appendix-c}{%
\subsection{Appendix C: System and Method
Details}\label{appendix-c}}

\hypertarget{app-telemetry}{%
\subsubsection{C.1 Telemetry collectors and store
schema}\label{app-telemetry}}

Three collectors provide the inference labels. A per-job emitter records
cold start, request and aggregate throughput, TTFT,
inter-token-latency percentiles, KV usage, and power for each serving
run. It also records the serving-window epochs. A per-job DCGM collector
aggregates power and activity. Its serving-window path restricts averages
to the recorded \texttt{{[}serve\_start,\ serve\_end{]}} interval to avoid
the staleness artifact in §2.3. A cluster-wide harvester reads all-user
jobs from Slurm \texttt{sacct} with our HPC\_Tools collection scripts
{[}28{]}. It decodes the per-job statistics blob stored in each job's
admin comment by the \texttt{jobstats} platform {[}29{]}, which is
present on roughly 92\% of jobs, and
stores users as salted hashes for privacy.

The safety shield consumes GPU-health signals, including XID errors and
row-remap failures. The personal store contains 543 job-statistics rows,
481 per-job DCGM rows, and 11 DCGM snapshots alongside the 561 inference
runs. Training and recovery collectors also exist in the shared
substrate, as do hardware-fingerprinting probes for node admission,
FLOP-utilization baselines, and checkpoint throughput. The inference
advisor does not consume those signals.

\hypertarget{app-updates}{%
\subsubsection{C.2 Gated model updates}\label{app-updates}}

\emph{The loop.} Every launch produces an inference-run row. Periodic
rebuilds backfill realized job statistics and DCGM data, refit a
candidate model, and gate that candidate against the incumbent before
promotion. This loop carries upstream improvements into production,
including the serving-window telemetry fix (§2.3) and new precision or
family data.

\emph{The no-regression gate.} The gate's decision target is aggregate
tokens/s, which drives placement utility. A candidate is promoted only if
its decision-target error does not regress by more than 0.5 percentage
points and it was trained on at least as many rows as the incumbent. The
first fit always promotes. This mirrors validation gates in production
ML platforms {[}16{]}, {[}17{]} and regression avoidance in continual
learning {[}18{]}, specialized here to a cost model rather than a
classifier.

\emph{Provenance and envelope co-evolution.} Each version is recorded in
a provenance ledger with its training-row count, decision-target
metrics, and support envelope. Any recommendation can therefore be traced
to a model version. The support envelope is recomputed at each fit, so
the abstention boundary changes only when data changes. In the observed
run, incorporating holdout data expanded the context bound to 32,768 and
the concurrency bound to 2,000. This single promotion cycle moved the
dataset from 447 to 460 rows, with MAPE moving from 5.89\% to 6.19\%.

\hypertarget{app-recal}{%
\subsubsection{C.3 Held-out-cell recalibration and the nested
confirmation}\label{app-recal}}

The offset is the conformal quantile of the grouped-CV held-out-cell
residual pool. It reaches the 80\% target on that pool by construction
and, under exchangeability, on future cells drawn from the same
distribution. The measured quantities are therefore the baseline
under-coverage, 64--69\%, and the width cost of recalibration,
$\times$1.04--1.12 (Table~\ref{tab:recal}). The offset is computed once
from grouped cross-validation and stored with the model, so deployed
intervals use held-out-cell-calibrated widths.

We also test the offset with a nested grouped cross-conformal split. In
this check, the offset is sized on workload cells that are \emph{disjoint}
from the cells used to measure coverage. The random-fold offset
under-covers again: 66\% for aggregate throughput, 73\% for TTFT, and
67\% for KV-cache usage. The held-out-cell offset restores coverage to
80--84\%, with a larger width cost of $\times$1.34--1.51. Because the
number of cells is small (56--98), these nested estimates have wide
bootstrap confidence intervals over cells.

\hypertarget{app-voi}{%
\subsubsection{C.4 Benchmark-gate internals}\label{app-voi}}

The value of a benchmark \texttt{b} is the expected improvement in the
best achievable utility after observing \texttt{b}. Equivalently, it is
the expected best utility after the probe minus the best utility under
current uncertainty. We estimate this quantity with Monte Carlo sampling
using $K=32$ draws from the model's predictive distribution. The gate
then probes sequentially with a 0.2 node-hour budget, depth 3, and
epsilon 0.05. It stops on negative net value, sub-epsilon improvement,
budget exhaustion, or exhaustion of the six-benchmark menu. Each
benchmark has a closed-form node-hour cost. This is related to
efficient-evaluation work {[}12{]}, but the question differs: whether to
spend node-hours measuring before launch, not how many examples are
needed to score a model.

We tested the gate across eight scenarios spanning cost-conscious,
throughput-first, and SLO-constrained operating points. The scenarios
include close, high-uncertainty arms, such as H100 versus H200 for a 7B
model at concurrency 1000. In that case, predicted throughputs differ by
a few percent and intervals span $\sim$45\% of the median. The v1
\texttt{condition\_on(outcome)} shrinks the affected target's predictive
interval by 0.4$\times$ but leaves the central estimate unchanged. For
one arm, a sampled throughput outcome of 37,009 tok/s left the q50 at
29,849 tok/s while narrowing q10--q90 width from $\sim$14,200 to
$\sim$5,700 tok/s. The conditioned utility therefore equals the
unconditioned utility, and the Monte Carlo value of information is zero
by construction. A useful fix must refit on augmented data per sample, or
otherwise update the posterior mean. The current point-quantile GBMs do
not support that update cheaply.

\end{document}